\documentclass[aps,prl,twocolumn,groupedaddress,preprintnumbers]{revtex4}

\def\mylimit#1{\mathrel{\mathop{\kern0pt\longrightarrow}\limits_{#1}}}

\makeatletter
\@addtoreset{equation}{section}
\makeatother

\newcommand{\VEV}[1]{\left\langle #1 \right\rangle}

\newcommand{\nn}{\nonumber}

\newcommand{\bequ}{\begin{equation}}
\newcommand{\eequ}{\end{equation}}
\newcommand{\beqn}{\begin{eqnarray}}
\newcommand{\eeqn}{\end{eqnarray}}
\newcommand{\bctr}{\begin{center}}
\newcommand{\ectr}{\end{center}}

\begin{document}
\preprint{\vbox{\hbox{hep-ph/0212141}
\hbox{KUNS-1816}
\hbox{\today}}}

\title{Non-Abelian Horizontal Symmetry and Anomalous $U(1)$ Symmetry
for the Supersymmetric Flavor Problem
}

\author{Nobuhiro {\sc Maekawa}}
\email[]{maekawa@gauge.scphys.kyoto-u.ac.jp}
\affiliation{Department of Physics, Kyoto University, Kyoto 606-8502, Japan}

\date{\today}

\begin{abstract}
It is shown that using non-abelian horizontal gauge symmetry and anomalous
$U(1)_A$ symmetry in grand unified theories (GUTs),
realistic quark and lepton mass matrices 
including large neutrino mixings can be obtained, while
the differences among the scalar fermion 
 masses are sufficiently small
for suppression of various 
flavor changing neutral current (FCNC) processes, especially in $E_6$ GUT.
Combining the Higgs sector, in which doublet-triplet splitting is realized,
a complete $E_6\times SU(3)_H$ GUT, in which 
three generations
are unified into a single multiplet, $\Psi({\bf 27},{\bf 3})$, is obtained.

\end{abstract}


\maketitle


Two fundamental features, that have not yet been understood, of the standard 
model (SM) are the hierarchical structure of the Yukawa couplings and
the replication of quarks and leptons. GUT provides a description of
the unification of 
one family of quarks and leptons, but
offers no understanding their generations. 
Introducing a horizontal symmetry is a natural way to distinguish 
these generations
and to realize the hierarchical structure of the Yukawa couplings.
Such an approach has been studied 
in the literature, with abelian horizontal symmetry \cite{abel,TGUT,BM}, 
non-abelian horizontal
symmetry \cite{nonabel}, and discrete symmetry.
Once supersymmetry (SUSY) is introduced
to stabilize the weak scale, 
it is, in many cases, necessary that the first and second generation 
scalar fermion (sfermion) masses
be nearly identical to suppress the FCNC processes.
In SUSY theories, one of the most important problems is to satisfy
the above two antithetical aspects of flavor physics (SUSY flavor problem).
It is quite interesting that by introducing one of the various kinds of
horizontal symmetries, this SUSY flavor problem can potentially be solved
\cite{Sabel,Snonabel,Sdiscrete}.
Anomalous $U(1)_A$ gauge symmetry \cite{U(1)}, an abelian horizontal symmetry,
whose anomaly is cancelled by the Green-Schwarz mechanism \cite{GS},
 can accomplish this, though the artificial structure of Yukawa matrices
 must be assumed \cite{Sabel}.
A non-abelian horizontal symmetry may be more interesting, because 
it is obvious that this type of symmetry results in the degeneracy of 
sfermion masses,
though it is not easy to obtain realistic quark and lepton
mass matrices in a simple way, keeping the suppression of the FCNC processes 
\cite{Snonabel}.
In this paper, we show that using both such symmetries,
the SUSY flavor problem is naturally solved,
in particular, in the GUT scenario proposed in Refs. \cite{TGUT,BM,Unif,MY}.

It is worthwhile recalling the basic features of non-abelian horizontal 
symmetry, and for this purpose, we consider a simple model with 
horizontal symmetry $U(2)$.
If under  $U(2)$ the three generations of quarks and leptons,
$\Psi_i=(\Psi_a,\Psi_3)$ ($a=1,2$), transform as ${\bf 2+1}$, and
the Higgs field $H$ is a singlet, then only the Yukawa couplings for 
the third generation are allowed by the horizontal symmetry.
This accounts for the large top Yukawa coupling. 
The $U(2)$ horizontal symmetry is broken by the two vacuum expectation values 
(VEVs) of the doublet $\VEV{\bar F^a}=\delta^a_2V$ and of the anti-symmetric
tensor $\VEV{A^{ab}}=\epsilon^{ab}v$ ($\epsilon^{12}=-\epsilon^{21}=1$) as
\begin{equation}
  U(2)_H \mylimit{V}U(1)_H \mylimit{v} {\rm nothing}.
\end{equation}
The ratios of the VEVs to the cutoff, $\epsilon\equiv V/\Lambda\gg
\epsilon'\equiv v/\Lambda$, yield the following hierarchical structure of
the Yukawa couplings:
\begin{equation}
Y_{u,d,e}\sim\left( \matrix{ 0 & \epsilon' & 0 \cr
               \epsilon' & 0 & \epsilon \cr
               0 & \epsilon & 1}\right).
\end{equation}
Moreover,
the $U(2)_H$ symmetric interaction
 $\int d^4\theta \Psi^{\dagger a}\Psi_a Z^\dagger Z$, where $Z$ 
 has a non-vanishing VEV given by $\VEV{Z}\sim \theta^2\tilde m$,
leads to nearly equal first and second generation sfermion masses,
with
\begin{equation}
\tilde m^2_{u,d,e}\sim \tilde m^2\left(\matrix{1 & 0 & 0 \cr
                                      0 & 1+\epsilon^2 & \epsilon \cr
                                      0 & \epsilon & O(1) }\right),
\end{equation}
where the difference between these masses, $\epsilon^2$, results
from higher
dimensional interactions, like 
$\int d^4\theta(\Psi_a\bar F^a)^\dagger \Psi_b\bar F^bZ^\dagger Z$,
through a non-vanishing VEV $\VEV{\bar F}$. 
These mass matrices lead to the relations
$
\frac{\tilde m_2^2-\tilde m_1^2}{\tilde m^2}
\sim \frac{m_{F2}}{m_{F3}},
$
where $m_{Fi}$ and $\tilde m_i$ are the masses of the $i$-th generation fermions
and the $i$-th generation sfermions, respectively.
Unfortunately, these predictions of this simple model imply
a problematic contribution to the
$\epsilon_K$ parameter in $K$ meson mixing and 
the $\mu\rightarrow e\gamma$ process.
Moreover, it is obvious that hierarchical
Yukawa couplings predicted by this simple model are similar for the 
up-quark sector, the down-quark sector, and the lepton-sector.
This is inconsistent with experimental results.
In particular, in neutrino sector, with this model
it seems to be difficult to
obtain the large neutrino mixing angles that have been measured in some recent 
experiments \cite{atmos,solar}. 
Several models in which some of these problems can be avoided have been
studied in the literature
\cite{Snonabel}, but there is no existing formulation in which all of these
problems can be avoided in a natural manner.

In this paper, we consider a new approach employing anomalous $U(1)_A$ gauge
symmetry. We show that this gauge symmetry allows for all of these problems
to be solved in a natural manner.
The author and collaborators
have already pointed out that employing the anomalous $U(1)_A$ gauge symmetry 
allow us to solve various problems that plague GUTs
\cite{TGUT,BM,Unif,MY}, for example, the doublet-triplet splitting
problem, proton instability, unrealistic GUT relations between quark and lepton
Yukawa matrices, and unnatural gauge coupling unification. 
One of the most important features of the GUT scenario is that 
the theory can be defined once we fix the anomalous $U(1)_A$ charges, because
generic interactions are introduced.
Vacuum expectation values (VEVs) are determined
by anomalous $U(1)_A$ charges as
\begin{equation}
\VEV{O_i}\sim \left\{ 
\begin{array}{ccl}
  \lambda^{-o_i} & \quad & o_i\leq 0 \\
  0              & \quad & o_i>0
\end{array} \right. ,
\label{VEV}
\end{equation}
where the $O_i$ are GUT gauge singlet operators with 
charges
$o_i$, and $\lambda\equiv \VEV{\Theta}/\Lambda\ll 1$.
Here the Froggatt-Nielsen (FN) field $\Theta$ has 
an anomalous $U(1)_A$ charge of $-1$ \cite{FN}. 
(In this paper we choose $\Lambda\sim 2\times 10^{16}$ GeV, which results
from the natural gauge coupling unification \cite{Unif}, 
and $\lambda\sim 0.22$.)
Throughout this paper, we denote all superfields and chiral 
operators by uppercase letters and their anomalous $U(1)_A$ 
charges by the corresponding lowercase letters. When convenient, we
use units in which $\Lambda=1$. 
 Such a vacuum structure is naturally obtained if we introduce generic
interactions even for higher-dimensional operators and if the $F$-flatness 
conditions
determine the scale of the VEVs. In this paper, we show that
by applying the vacuum relation (\ref{VEV}) to the Higgs which breaks
the non-abelian horizontal gauge symmetry, 
realistic quark and lepton mass matrices including large neutrino mixing 
angles can be obtained, while the FCNC processes are suppressed.
In other words, we show how the FCNC processes can be 
suppressed
by introducing non-abelian horizontal gauge symmetry into our GUT scenario,
in which 
realistic Yukawa mass matrices have already been obtained.
We should note that horizontal gauge symmetry may introduce a problem,
because the 
non-vanishing $D$-term may break the degeneracy of the sfermion masses.
Therefore it may necessary to include some mechanism that suppresses 
the $D$-term, as in 
Ref. \cite{Babu}, but in this paper, we do not discuss this problem.

Let us explain the basic idea with an $SU(5)$ GUT model with 
$SU(2)_H\times U(1)_A$.
The field content is given in Table I.
\begin{center}
Table I. Typical values of anomalous $U(1)_A$ charges.
The half integer charges play the same role as R-parity.

\begin{tabular}{|c|c|c|c|c|c|c||c|c|c|c|c|c|} 
\hline
   &   $\Psi_a$ & $\Psi_3$ & $T_a$ & $T_3$ & $N_a$ &$ N_3$ & $H$  & $\bar H$ 
   & $F_a$ & $\bar F^a$ & $S$ & $\Theta$ \\
\hline 
 $SU(5)$ & {\bf 10} & {\bf 10} & ${\bf \bar 5}$ & ${\bf \bar 5}$ & {\bf 1}& 
 {\bf 1} & ${\bf 5}$ 
 & ${\bf \bar 5}$ & {\bf 1} & {\bf 1} &{\bf 1}& {\bf 1}  \\
 $SU(2)_H$ & {\bf 2} &{\bf 1} & {\bf 2} &{\bf 1} &{\bf 2} &{\bf 1} &
 ${\bf 1}$ & {\bf 1} &{\bf 2} &${\bf \bar 2}$ &{\bf 1}&{\bf 1}  \\
 $U(1)_A$ & $\frac{13}{2}$ & $\frac{7}{2}$ & $\frac{13}{2}$ & $\frac{11}{2}$ 
 & $\frac{13}{2}$ & $\frac{7}{2}$ & $-7$ & $-7$ & 
 $-2$ & $-3$ & 5 & $-1$  \\
 \hline
\end{tabular}
\end{center}
The superpotential for the $S$ field,
$
W_S=\lambda^s S(1+\lambda^{f+\bar f}\bar FF),
$
leads to the $SU(2)_H$ breaking VEV 
$
\VEV{\bar FF}\sim \lambda^{-(f+\bar f)}.
$
Without loss of generality, we can take
$
|\VEV{\bar F^a}|=|\VEV{F_a}|\sim \delta_{a2}\lambda^{-\frac{1}{2}(f+\bar f)},
$
using the $SU(2)_H$ gauge symmetry and its $D$-flatness condition.
Then, from the relations
\begin{equation}
\lambda^{\psi+\bar f}\Psi_a\VEV{\bar F^a}\sim 
\lambda^{\psi+\Delta f}\Psi_2, \ 
\lambda^{\psi+f}\epsilon^{ab}\Psi_a\VEV{F^b}\sim 
\lambda^{\psi-\Delta f}\Psi_1, \nn
\end{equation}
where $\Delta f\equiv \frac{1}{2}(\bar f-f)$,
it is obvious that with the effective charges defined as
$\tilde x_3\equiv x_3$, $\tilde x_2\equiv x+\Delta f$, 
and $\tilde x_1\equiv x-\Delta f$ for $x=\psi,t,n$,
the Yukawa matrices of the quarks and leptons $Y_{u,d,e,\nu}$ and 
the right-handed neutrino mass matrix $M_{\nu R}$ can be approximated as
\begin{eqnarray}
&&(Y_u)_{ij}\sim \lambda^{\tilde \psi_i+\tilde \psi_j+h},\
(Y_d)_{ij}\sim (Y_e^T)_{ij}\sim \lambda^{\tilde \psi_i+\tilde t_j+\bar h} \\
&&(Y_\nu)_{ij}\sim \lambda^{\tilde t_i+\tilde n_j+h}, \
(M_{\nu R})_{ij}\sim \lambda^{\tilde n_i+\tilde n_j} 
\end{eqnarray}
from the generic 
interactions 
$
W_{\rm fermion}=\tilde \Psi^2\lambda^hH+\tilde \Psi\tilde T
\lambda^{\bar h}\bar H+
\tilde T\tilde N\lambda^hH+\tilde N\tilde N,
$
where 
$\tilde X\equiv \lambda^{x+f}\epsilon^{ab}X_aF_b+\lambda^{x+\bar f}X_a
\bar F^a+\lambda^{x_3}X_3$ for $X=\Psi,T,N$.
Throughout this paper, we omit $O(1)$ coefficients for simplicity.
Then, the  neutrino mass matrix is obtained as
$
(M_{\nu})_{ij}= (Y_\nu)(M_{\nu R})^{-1}(Y_\nu^T)\frac{\VEV{H}^2}{\Lambda}
\sim\lambda^{\tilde t_i+\tilde t_j+2h}\frac{\VEV{H}^2}{\Lambda}.
$
In theories in which Yukawa couplings are determined by  $U(1)$ charges,
as in the above,  the unitary matrices $V_{y_P}$ $(y=u,d,e,\nu$ and 
$P=L,R)$ that diagonalize these 
Yukawa and mass matrices as  
$V_{y_L}^{\dagger}Y_yV_{y_R}=Y_y^{\rm diag}$, the Cabibbo-Kobayashi-Maskawa
matrix $V_{CKM}\equiv V_{d_L}V_{u_L}^\dagger$, and the Maki-Nakagawa-Sakata 
matrix $V_{MNS} \equiv V_{e_L}V_{\nu_L}^\dagger$
are roughly approximated by the matrices 
$(V_{\bf 10})_{ij}\equiv \lambda^{\tilde \psi_i-\tilde \psi_j}$ and
$(V_{\bf \bar 5})_{ij}\equiv \lambda^{\tilde t_i-\tilde t_j}$
as 
$V_{\bf 10}\sim V_{u_L}\sim V_{d_L}\sim V_{u_R} \sim V_{e_R}\sim V_{CKM}$
and
$V_{\bf \bar 5}\sim V_{d_R}\sim V_{e_L}\sim V_{\nu_L}\sim V_{MNS}$.
Using the typical charge assignment given in Table I, we obtain realistic 
structure of quark and lepton mass matrices, in which large neutrino mixing
angles are also realized.

The sfermion mass-squared matrices are written
\begin{equation}
\tilde m_y^2=\left(\matrix{\tilde m_{y_L}^2 & A_y^\dagger \cr
                           A_y & \tilde m_{y_R}^2 }\right).
\end{equation}
In this paper, we concentrate on mass mixings through $\tilde m_{y_P}^2$,
because a reasonable assumption like SUSY breaking in the hidden sector, 
leads to an $A_y$ that is proportional to the Yukawa matrix $Y_y$ \cite{SW}.
Roughly speaking, the sfermion mass squared matrix is given by
$\tilde m_{y_P}^2\sim \tilde m^2 {\rm diag} (1,1,O(1))$, and 
the correction 
$\Delta_{y_P}\equiv(\tilde m_{y_P}^2-\tilde m^2)/(\tilde m_{y_P})^2$ 
in the model described by Table I is approximately given by 
\begin{equation}
\Delta_{\bf 10}=\left(
\matrix{\lambda^5 & \lambda^6 & \lambda^{3.5}\cr
 \lambda^6 &\lambda^5 &  \lambda^{2.5} \cr
 \lambda^{3.5} & \lambda^{2.5}& R_{\bf 10} \cr }
\right),
\Delta_{\bf \bar 5}=\left(
\matrix{\lambda^5 & \lambda^6 & \lambda^{3.5} \cr
 \lambda^6 & \lambda^5 & \lambda^{4.5} \cr
 \lambda^{3.5} &  \lambda^{4.5}& R_{\bf \bar 5} \cr }
\right)
\end{equation}
for {\bf 10} fields and ${\bf \bar 5}$ fields.
Here $R_{\bf 10, \bar 5}\sim O(1)$.
For example, $(\Delta_{\bf \bar 5})_{12}$ can be derived using
the interaction 
$\int d^4\theta \lambda^{|f-\bar f|}(T\bar F)^\dagger (TF)Z^\dagger Z$.
Note that 
$(\tilde m_{d2}^2-\tilde m_{d1}^2)/\tilde m_d^2\sim 
(m_s/m_b)^2$, and the rather large neutrino
mixing angle $(V_{MNS})_{23}\sim \lambda^{0.5}$ can be realized in 
the model described by Table I.
The essential points are that the Yukawa hierarchy is determined by
the (effective) anomalous $U(1)_A$ charges, while
the corrections to the sfermion masses are determined by
the VEVs as $|\VEV{F}|=|\VEV{\bar F}|\sim \lambda^{-(f+\bar f)}$. 
Let us concentrate on the components $(Y_{d})_{32}$, $(Y_{e})_{23}$ 
(note $(Y_{d})_{32}\sim (Y_{e})_{23}$), and
 $(Y_{d,e})_{33}$,
which are required to be of the same order to obtain large atmospheric 
neutrino mixing.
The components $(Y_{e})_{23}$ and $(Y_{d})_{32}$ are
 obtained from the interaction 
$\Psi_3T_a \bar F^a\bar H$,
 and actually they are suppressed by the VEV  
 $\VEV{\bar F}\sim \lambda^{-\frac{1}{2}(f+\bar f)}=\lambda^{2.5}$. 
On the other hand, the components $(Y_{d,e})_{33}$ are also suppressed by
the factor $\lambda^{\psi_3+t_3+\bar h}\sim \lambda^2$, because of
the $U(1)_A$ symmetry.  It is obvious that smaller $(Y_{d,e})_{33}$
(namely, smaller $\tan\beta\equiv \VEV{H}/\VEV{\bar H}$)
allows for smaller $(Y_{e})_{23}$ and $(Y_{d})_{32}$, and therefore, 
smaller $\VEV{\bar F}$. This, then, leads to smaller corrections to the
sfermion masses.
In other words, because the  negative value of
the charge $\bar f$ can increase the Yukawa couplings  
$(Y_{e})_{23}$ and $(Y_{d})_{32}$
from the simply expected value 
$\lambda^{\psi_3+t+\bar h}=\lambda^2$ by a factor 
of $\lambda^{\bar f}=\lambda^{-2}$, 
larger values of these Yukawa couplings can be 
realized by smaller
VEVs of $F$ and $\bar F$, which lead to smaller corrections to the
sfermion masses.
Unfortunately, even with these smaller corrections, the FCNC processes
are not suppressed for two reasons, because the neutrino mixing angles are
large and because $R_{\bar 5}\sim O(1)$.
The various FCNC processes constrain the mixing matrices defined by
$\delta_{y_P}\equiv V_{y_P}^\dagger \Delta_{y_P}V_{y_P}$
\cite{GGMS}.
In this model, these
mixing matrices are approximated as  
\begin{equation}
\delta_{\bf 10}=\left(\matrix{\lambda^5 & \lambda^6 & \lambda^{3.5} \cr
                          \lambda^6 & \lambda^5 & \lambda^{2.5} \cr
                          \lambda^{3.5} & \lambda^{2.5} & R_{\bf 10} }
             \right), \ 
\delta_{\bf \bar 5}=R_{\bf \bar 5}
          \left(\matrix{\lambda^3 & \lambda^2 & \lambda^{1.5} \cr
                          \lambda^2 & \lambda & \lambda^{0.5} \cr
                          \lambda^{1.5} & \lambda^{0.5} & 1 }
             \right).
\label{delta}
\end{equation}
In order to suppress the contribution
to $\epsilon_K$ in $K$ meson mixing, scalar quark masses larger than
1 TeV  are required, 
and in order to suppress the $\mu\rightarrow e\gamma$ process, scalar lepton
masses larger than 300 GeV are required.

It is notewhorthy that in $E_6$ GUT with anomalous $U(1)_A$ symmetry,
$R_{\bf \bar 5}$ can be small enough to suppress the FCNC processes.
To understand this, first note that under $E_6\supset SO(10)\supset SU(5)$, 
the fundamental representation
${\bf 27}$ is divided as
\begin{equation}
{\bf27} \rightarrow {\bf 16}[ {\bf 10} +{\bf \bar 5}
+{\bf 1}]
+{\bf 10}[{\bf \bar 5'}+{\bf 5}]
+{\bf 1}[{\bf 1}].
\end{equation}
We introduce two pairs of ${\bf 27}$ and ${\bf \overline{27}}$ to break
$E_6$ into $SU(5)$. The VEVs 
$|\VEV{\Phi}|=|\VEV{\bar \Phi}|\sim \lambda^{-\frac{1}{2}(\phi+\bar \phi)}$
 break $E_6$
into $SO(10)$, which is broken into $SU(5)$ by the VEVs 
$|\VEV{C}|=|\VEV{\bar C}|\sim \lambda^{-\frac{1}{2}(c+\bar c)}$.
Because the three fundamental representation fields 
$\Psi_i({\bf 27})$ ($i=1,2,3$) include
 $3\times ({\bf 10}+{\bf 5})+6\times {\bf \bar 5}$ of $SU(5)$,
only three of the six ${\bf \bar 5}$ become massless.
The $3\times 6$ mass matrix is obtained from the 
interactions 
$W=\lambda^{\psi_i+\psi_j+\phi}\Psi_i\Psi_j\Phi
+\lambda^{\psi_i+\psi_j+c}\Psi_i\Psi_jC$. 
Note that $\psi_3<\psi_1,\psi_2$ because top quark has larger Yukawa 
couplings than the first and second generation fields.
Therefore, as discussed in Ref. \cite{BM}, 
it is natural that
these three massless ${\bf \bar 5}$ fields come from the first and second 
generation fields, $\Psi_1$ and $\Psi_2$, 
because the smaller charge $\psi_3$ results in larger masses of the 
$\bf \bar 5$ fields from $\Psi_3$.
If the first two multiplets become the doublet $\Psi({\bf 27}, {\bf 2})$ under
$SU(2)_H$ in this $E_6$ GUT, 
then it is obvious that the sfermion masses for these three modes
 ${\bf \bar 5}$ are equal at leading order.
Then, if we fix the model by setting $(f,\bar f)=(-2,-3)$ and
$(\psi,\psi_3,\phi,\bar \phi,c,\bar c)=(5,2,-4,2,-5,-2)$ (noting
that odd R-parity
is required for the matter fields $\Psi$ and $\Psi_3$), 
the massless modes become 
$({\bf \bar 5_1}, {\bf \bar 5_2}, {\bf \bar 5'_1}
+\lambda^\Delta{\bf \bar 5_3})$, where
$\Delta=\tilde \psi_1-\tilde \psi_3+\frac{1}{2}(\phi-\bar \phi-c+\bar c)=2$.
As discussed in Ref. \cite{MY}, it is natural that the Higgs fields $H$ and 
$\bar H$ are included in ${\bf 10_{\Phi}}$ of $SO(10)$. Therefore, the 
${\bf \bar 5'}$ fields have no direct Yukawa couplings with $\bar H$.
The massless mode $ {\bf \bar 5'_1}+\lambda^\Delta{\bf \bar 5_3}$ has
Yukawa couplings only through the mixing with ${\bf \bar 5_3}$. 
Then the structure of the quark and lepton Yukawa matrices
becomes the same as that found in the previous $SU(5)$ model.
The correction to the sfermion masses  $\delta \tilde m_{\bf \bar 5}$ 
can be approximated from the
higher dimensional interactions as
\begin{equation}
\frac{\delta \tilde m_{\bf \bar 5}^2}{\tilde m^2}
\sim \left(
\matrix{\lambda^5 & \lambda^6 & \lambda^{5.5} \cr
 \lambda^6 & \lambda^5 &  \lambda^{4.5} \cr
 \lambda^{5.5} & \lambda^{4.5}& \lambda^2  }
\right),
\end{equation}
which leads to the same $\delta_{\bf \bar 5}$ as that in Eq. (\ref{delta})
if we use $R_{\bf\bar 5}=\lambda^2$. This decreases the lower limit
of the scalar quark mass to an acceptable value, 250 GeV. 
Note that $R_{\bar 5}$ can be obtained
as
$R_{\bf \bar 5}=(\delta \tilde m_{\bf \bar 5}^2/\tilde m^2)_{33}\sim 
\VEV{\bar \Phi\Phi}\sim \lambda^{-(\phi+\bar \phi)}$
from the interaction 
$\int d^4\theta \Psi^\dagger \Phi^\dagger \Psi\Phi Z^\dagger Z$.

Another interesting feature of $E_6$ GUT with anomalous $U(1)_A$ symmetry
is that we can extend the horizontal gauge group
to $SU(3)_H$. In this model the three generations of quarks and leptons
can be unified into a single multiplet, $\Psi({\bf 27},{\bf 3})$.
Supposing that the horizontal gauge symmetry $SU(3)_H$ is broken 
by the VEVs of two pairs of Higgs fields $F_i({\bf 1},{\bf 3})$ and 
$\bar F_i({\bf 1},{\bf \bar 3})$ $(i=2,3)$ as
$
|\VEV{F_{ia}}|=|\VEV{\bar F_i^a}|\sim 
\delta_i^a\lambda^{-\frac{1}{2}(f_i+\bar f_i)},
$
the effective charges can be defined from the relations 
\begin{eqnarray}
&&\lambda^{\psi+\bar f_i}\Psi_a\VEV{\bar F_i^a}\sim 
\lambda^{\psi+\frac{1}{2}(\bar f_i-f_i)}\Psi_i (i=2,3), \\
&&\lambda^{\psi+f_2+f_3}\epsilon^{abc}\Psi_a\VEV{F_{2b}F_{3c}}\sim 
\lambda^{\psi-\frac{1}{2}(\bar f_2-f_2+\bar f_3-f_3)}\Psi_1, \nn
\end{eqnarray}
as
$\tilde \psi_i\equiv \psi+\frac{1}{2}(\bar f_i-f_i),\quad
\tilde \psi_1\equiv \psi-\frac{1}{2}(\bar f_2-f_2+\bar f_3-f_3).
$
Then, if we choose their charges as $(f_3,\bar f_3,f_2,\bar f_2)=(2,-2,-3,-2)$ 
and $\psi=4$, this $E_6\times SU(3)_H$ model gives the same predictions
for the mass matrices of fermions and sfermions as the previous
$E_6\times SU(2)_H$ model. [The model obtained by choosing  
$(f_3,\bar f_3,f_2,\bar f_2)=(2,-3,-4,-3)$, $\psi=13/2$, and 
$(\phi,\bar \phi,c,\bar c)=(-7,3,-8,0)$ may be more interesting, because
mass matrices for quarks and leptons that are essentially the same
as those in Ref. \cite{BM}
are obtained if we set $\lambda^{1.5}=0.22$. ]

For both models $E_6\times SU(2)_H$ and $E_6\times SU(3)_H$, if we add
a Higgs sector that breaks $E_6$ into the gauge group of the standard model,
as in Ref. \cite{MY},  then we can obtain complete
$E_6\times SU(2)_H$ and $E_6\times SU(2)_H$ GUT, in which 
the degeneracy of the sfermion masses is naturally obtained.
As discussed in Refs. \cite{BM,MY}, these models yield not only 
realistic quark and
lepton mass matrices but also natural doublet-triplet splitting.
As an example, the entire Higgs content, except singlets under $E_6$, 
is listed in Table II.
\begin{center}
Table II. Typical values of anomalous $U(1)_A$ charges.

\begin{tabular}{|c|c|c|c|c|c|c|c|c|c|} 
\hline
   &    $A$ & $A'$ & 
   $\Phi$ & $\bar \Phi$ & $C$ & $\bar C$  & $C'$ &
   $\bar C'$ \\
\hline 
 $E_6$  &{\bf 78} & {\bf 78} 
 & {\bf 27} & ${\bf \overline{27}}$ &{\bf 27} & ${\bf \overline{27}}$ 
  &{\bf 27} & ${\bf \overline{27}}$ \\
 $U(1)_A$ & $-1$ & $5$ & 
 $-4$ & $2$ & $-5$ & $-2$ & $9$ & $7$  \\
 $Z_2$  & $-$ & $-$ & + & + & + & + & $-$ & $-$ \\
 \hline
\end{tabular}
\end{center}
Here, all the Higgs fields are singlets under the horizontal gauge 
symmetry $SU(2)_H$ or $SU(3)_H$. Therefore additional fields that are 
not singlets under the horizontal gauge symmetry are required for anomaly
cancellation, for example, a doublet for $SU(2)_H$ models and a 
${\bf\overline{10}}$
for $SU(3)_H$ models. It may be interesting to introduce
non-singlet Higgs fields under the horizontal gauge symmetry. 
This subject will be examined in the near future.

The $E_6\times SU(3)_H$ GUT and $E_6\times SU(2)_H$ GUT give interesting 
predictions for the structure of sfermion masses, in addition to 
the predictions on $B$-physics which are examined in
the literature \cite{Okada}. 
Roughly speaking,
 all the sfermion fields have nearly equal masses, except the third generation
 fields included in {\bf 10} of $SU(5)$.
More precisely, this degeneracy is lifted by $D$-term 
contributions of $SU(3)_H$ and $E_6$. 
Though the size of such contributions vary greatly among the models,
and some of these contributions
must be small in order to suppress the FCNC processes, it is important
to test these GUT models
with precisely measured masses of sfermions, as discussed 
in Ref. \cite{KMY}.

  N.M. is supported in part by Grants-in-Aid for Scientific 
Research from the Ministry of Education, Culture, Sports, Science 
and Technology of Japan.

\end{document}